\documentclass[runningheads,11pt]{article}

\pdfoutput=1 
\usepackage{lscape, rotating}  
\usepackage{amsfonts}
\usepackage[]{color}
\usepackage{amsfonts}
\usepackage{amsmath}
\usepackage{caption}
\usepackage{lastpage}
\usepackage[utf8]{inputenc}
\usepackage{graphicx}%
\usepackage{subfig}  
\usepackage{setspace}
\usepackage{fancyhdr}
\usepackage{color}
\usepackage[mathlines]{lineno}
\usepackage{lscape}
\usepackage[english]{babel}
\usepackage{epsfig}
\usepackage{float}	% fix figures positions
\usepackage{makecell}
\usepackage{multirow}		% multiple row cells
\usepackage{bbold}	% indicatrice
\usepackage{indentfirst}	% Identation after each heading section
\usepackage{bm} % Bold in math
\usepackage{arydshln}  % get dashed lines
\usepackage{cite}  % To compress consecutive references e.g. [3-7]

\usepackage{authblk}     % authors and affiliations
\usepackage{titlesec} % For formatting sections
\usepackage{lineno}    % Number lines%
\usepackage[bookmarks=true,colorlinks=true,linkcolor=blue,citecolor=blue]{hyperref}
\usepackage[toc,page,title,titletoc,header]{appendix}
\usepackage{appendix}
\usepackage{newfloat}
\usepackage{longtable}
\usepackage{multirow}		% multiple row cells
\usepackage{subfig}
\usepackage{pstricks}
\usepackage{geometry}
\usepackage{booktabs}
\usepackage{natbib}

\title{ A  right-truncated Poisson mixture model  for analyzing count data}

\author[$1,2^*$]{Babagnidé François \bf{KOLADJO}}
\author[$2$]{Ricardo Anderson \bf{DONTE}}
\author[$3$]{Epiphane \bf{SODJINOU}}

\affil[$1$]{\it {Ecole Nationale de Statistique, de Planification et de Démographie ENSPD, Université de Parakou, Parakou, Bénin}} \vspace{-1cm}
\affil[$2$]{\it {Laboratoire de Recherche en Science de Population et Développement (LaReSPD), Université de Parakou, Parakou, Bénin}}   \vspace{-1cm}

\affil[$3$]{\it {Faculté d'agronomie, Université de Parakou, Parakou, Bénin}}   \vspace{-1cm}

\affil[*]{Corresponding author: francois.koladjo@gmail.com}
\date{}

\newtheorem{theorem}{Theorem} 
\begin{document}

	\renewcommand{\abstractname}{Abstract}
	%	\linenumbers
	\maketitle
	\vspace{-1cm}
	\begin{abstract} 
\noindent In this paper, we investigate right-truncated count data models incorporating covariates into the parameters. A regression method is proposed to model right-truncated count data exhibiting high heterogeneity. The study encompasses the formulation of the proposed model, parameter estimation using an Expectation Maximisation  (EM) algorithm, and the properties of these estimators. We also discuss model selection procedures for the proposed method. Furthermore, a Monte Carlo simulation study is presented to assess the performance of the proposed method and the model selection process. Results express accuracy under  regularity conditions of the model. The method is used to analyze the determinants of the degree of adherence to preventive measures during the COVID-19 pandemic in northern Benin. The results show that a  right-truncated Poisson mixture model is adequate to analyze these data. Using this model, we conclude that age,  education level, and household size determine an individual's degree of adherence to preventive measures during COVID-19 in this region.
		\medskip
		\end{abstract}

\section{Introduction}\label{sec1}

 The Poisson regression model is a statistical method designed to explain a count variable using a set of covariates. This model is relies on the assumption of equidispersion.  In other words, the method assumes equality between the mean and variance of the count variable. Applications to real-world data have revealed that this assumption of equidispersion is not always met. In practice, the dispersion of counts often exceeds what would be expected in a standard Poisson model ~\citep{breslow1984extra}. The authors investigated this phenomenon for count data.  For the classical (Unicomponent) Poisson regression model, we found that  overdispersion affects both the model fit and estimates. The inadequacy of the unicomponent Poisson regression model in the presence of overdispersed count data has motivated extensive research in the development of alternative models capable of effectively accounting for the  characteristics of these data. Some notable contributions include negative binomial regression \citep{hausman1984econometric}, quasi-likelihood-based models ~\citep{williams1982extra, breslow1984extra}, random effects models \citep{schall1991estimation}, and mixture models \citep{manton1981variance,hinde1982compound,follmann1989generalizing}. Mixture models are well positioned for analyzing overdispersed data, as they are able to account for overdispersion related to population heterogeneity.
   In a regression context, these models assume that each observation comes from an unobserved component of the population, with each component characterized by its own set of regression coefficients. For example, \cite{wang1996mixed} proposed a regression method with a mixture of Poisson distributions and used it to analyze  count data affected by overdispersion.

However, few studies have considered mixture regression models for right-truncated count data. As a consequence of truncation, neither the classical Poisson distribution,  nor a mixture of Poisson distributions are appropriate. The regression methods for count data mentioned above may also be inadequate for explaining truncated counts using  covariates. This motivated our investigation of the unicomponent and  mixture of right-truncated Poisson distributions.

In section 2, we present  a brief introduction to right-truncated Poisson regression  and right-truncated Poisson mixture models. Section 3 addresses the maximum likelihood estimation method, the properties of the estimators, an algorithm for the parameter estimation,  a method for selecting the number of mixture components,  and overdispersion testing for right-truncated counts models. In Section 4, we present and  discuss results from Monte Carlo simulations evaluating the method's accuracy. Section 5 applies the proposed method to adherence data for preventive measures during COVID-19, whereas Section 6 discusses the findings.

\section{Regression models for right-truncated count data} \label{sec2}

\subsection{Right-truncated Poisson regression  model}\label{subsec1}

The right-truncated Poisson regression model $($the RTPR model$)$ is a statistical approach used to model a right-truncated count variable given a set of covariates. It assumes that the counts are independently and identically distributed according to a right-truncated Poisson distribution.
We say that a random variable $Y_i $ is distributed according to a $\tau$-right-truncated Poisson distribution with mean $\lambda $ if its probability mass function (p.m.f) at $y_i$ is given by :
\begin{equation}
	P(Y_i = y_i) = \frac{\frac{\lambda^{y_i} e^{-\lambda}}{y_i!}}{\sum \limits_{k=0}^{\tau} \frac{\lambda^{k}e^{-\lambda}}{k!}} \label{eq1}
\end{equation}
The expression in equation \eqref{eq1} represents the probability that $Y_i$ takes the non-negative integer value  $y_i \in \left\{ 0,1,2,...,\tau \right\}$  with $\tau \geq 2 $  the truncation  threshold, and  $\lambda$ being  the mean parameter of a Poisson distribution. We consider a non-severe truncation such that $ \tau$ is not less than $\lambda.$ 
A log-linear specification is used, analogous to the  link function in the classical Poisson regression model. That is, the linear predictor is defined as 
\begin{equation}
log(\lambda_{i}) = \sum \limits_{k=0}^{K} \beta_{k}x_{ik}, \quad \beta_k \in  \mathbf{R}. \label{eq2}
\end{equation}
where $x_i = \left(x_{i0},x_{i1},...,x_{iK}\right)$ is the vector   of $K$ explanatory variables $($with intercept term $x_{i0})$ measured for individual $i.$ The  $ \beta_k $ denote the regression coefficients. The  RTPR model differs from the Poisson regression model in two important ways. Unlike the latter, it allows for under-dispersion, as stated in Theorem~\ref{eq1} below.  On the other hand, when the truncation threshold $\tau$  in equation~\eqref{eq1}  grows to infinity, the RTPR becomes the classical Poisson regression model. In this case, it can be viewed as a generalisation of the Poisson regression model to account for underdispersion in counts data analysis.

\smallskip

\begin{theorem} \label{theo1}
	Let $Y$ be a random variable distributed according to a  right-truncated Poisson distribution with threshold  $\tau$ with parameter $\lambda$ and p.m.f. as given  in equation~\eqref{eq1}.  Then we have:
	\begin{equation}
			\forall  \lambda  \leq  \tau,	V(Y) < E(Y) 
	\end{equation} \label{eq3}   
\end{theorem}

The proof of  Theorem~\ref{theo1}  is given in Appendices (subsection ~\ref{subsubsec8.1.1}).  From this  result, without loss of generality the RTPR model is adequate  for Poisson regression modeling where the response variable displays  underdispersion.  Hence, where heterogeneity due to overdispersion is the main concern, the RTPR model is no longer suitable, even when the data are right-truncated. These data characteristics are frequently observed. Below, we present  a mixture model to account for both heterogeneity and right-truncation. 

\subsection{Regression using a mixture of right-truncated Poisson distributions}\label{subsec2}
Let $\left(y_1,y_2,...,y_n\right)$ be an observation  of the random vector $\left(Y_1,Y_2,...,Y_n\right)$, where $Y_i$ represents the number of  events within a given time interval for  individual i. \textit{Assume that $Y_i$ has finite support points equal to the set $\left\{0,1,...,\tau\right\}$. }  Let  $X$ be a  $(nxK)$ full-rank matrix with i$^{th}$ row equal to the observed vector $x_i = (x_{i1},x_{i2},...,x_{iK})$ of $K$ explanatory variables on the i$^{th}$ individual in a sample of size $n.$ Under the heterogeneity assumption, the proposed  regression model, based on a mixture of right-truncated Poisson distributions,  assumes the following:

\begin{itemize}	
		\item For each observation $y_i$, there exists an unobserved discrete random variable $Z_i$ modeling the component or subpopulation from which observation $y_i$ is generated;
		\item  A priori, each individual $i$ can arise from any of $J$ components of the mixture;
		\item The random variables $(Y_i,Z_i)$, $i \in \left\{1,2,...,n\right\}$, are independent;
		\item For each $i$, $Z_i$ is distributed according to a multinomial  distribution with $J$ support points such that $\sum \limits_{j=1}^{J}p_j = 1$ ; where $p_j$ is the probability of the event $\left[Z_i = j\right]$;
		\item Conditional on the event $\left[Z_i = j\right],$  the random variable $Y_i$ is distributed according to a $\tau-$ right-truncated  Poisson distribution : 
		\begin{equation}
			P(Y_i = y_i|Z_i = j) = \frac{P_o(y_i,\lambda_{ij})}{\sum \limits_{k=0}^{\tau} P_o(y_i,\lambda_{ij})}  = \frac{  \frac{(\lambda_{ij})^{y_i} e^{-\lambda_{ij}}}{y_i!}}{  \sum \limits_{k = 0}^{\tau} \frac{(\lambda_{ij})^{k} e^{-\lambda_{ij}}}{k!}}  
		\end{equation}
		
		\item For each $i \in \left\{1,...,n\right\} $, 
		\begin{equation}
			ln(\lambda_{ij}) =  \beta_j^{'}x_i = \sum_{k=1}^{K}\beta_{jk}x_{ik} \label{eq5}
		\end{equation} 
		
		where $\beta_j = \left(\beta_{1j},\beta_{2j},...,\beta_{Kj}\right)$ is the        vector of regression coefficients for the $j^{th}$ component of the mixture. Like equation \eqref{eq2}, equation \eqref{eq5}  is a link function specifying the mean parameter $(\lambda)$ as  a function of covariates.  It guarantees strict   positivity of the Poisson mean parameter. \cite{mccullagh1989generalized}, \cite{lee1986specification}, \cite{wang1996mixed} are some examples of similar works.
\end{itemize}

Under the assumptions defined  above, the unconditional distribution of the variable $Y_i$ is a mixture of right-truncated Poisson distributions with constant mixing proportions. The conditional distribution of $Y_i$ given the $j^{th}$ component of the mixture is a $\tau$-right-truncated Poisson distribution with mean $\lambda_{ij}.$ Therefore, the probability density function of $Y_i$ is given by: 
		
		\begin{equation} 
			P(Y_i = y_i) = \sum_{j = 1}^{J} p_j \times \frac{P_o (y_i, \lambda_{ij})}{\sum \limits_{k = 0}^{\tau} P_o(k, \lambda_{ij})}\label{eq6}
		\end{equation}

The unconditional mean and variance, according to the above model, are respectively 
		
		\begin{equation}
			E(Y_i)  =  \sum \limits_{j=1}^{J} p_j \times \lambda_{ij}\frac{F_{\lambda_{ij}}(\tau-1)}{F_{\lambda_{ij}}(\tau)} \quad and 
		\end{equation}
		
		\begin{equation}
			V(Y_i) =  \sum \limits_{j=1}^{J}p_j \left(\lambda_{ij}^2\frac{F_{\lambda_{ij}}(\tau-2)}{F_{\lambda_{ij}}(\tau)} \right)  +  \sum_{j=1}^{J}\lambda_{ij}\frac{F_{\lambda_{ij}}(\tau-1)}{F_{\lambda_{ij}}(\tau)} - \left[\sum_{j=1}^{J} p_j\lambda_{ij} \frac{F_{\lambda_{ij}}(\tau-1)}{F_{\lambda_{ij}}(\tau)}\right]^2
		\end{equation}
Equations \eqref{eq5} and \eqref{eq6} define the regression model based on the mixture of right-truncated Poisson distributions.  This model accounts for heterogeneity in the presence of right truncation.

\section{Estimator derivation and their properties}\label{sec3}

The Expectation-Maximization (EM) algorithm proposed by~\cite{dempster1977maximum} is  a popular tool for parameter estimation  in mixture models with a fixed number of  components. We use an EM-type algorithm combined with a   quasi-Newton optimization method \citep{nash2018compact}  for parameter estimation in the mixture of  $\tau$-right-truncated Poisson distributions.  The steps of this algorithm are outlined below, after stating the estimating  equations  and the convergence properties of the estimators. 

\subsection{Estimating equations}\label{subsec3}

 Consider the observed data $\left(y,x\right)$  $\cong$ $\left(y_i,x_i\right),$ $i \in $ $\left\{1,2,\dots,n\right\}$, where the $y_i'$s denote the realizations of  a random variable $Y$ with p.m.f.  described by equation (\ref{eq6}).  The EM algorithm 
 treats an individual's membership in one of the $J$ mixture components as realizations of an unobservable random variable $Z.$ Let
  $\left(y,x,z\right)$ $\cong$ $\left(y_i,x_i,z_i\right)$ , $i$ $\in \left\{1;2;...;n\right\}$, denote the so-called  “incomplete” data with 
\begin{equation}		
	z_{ij} =  \begin{cases}
	1 & \text{if} \quad i \in C_j \\
	0 & \text{if} \quad i \notin C_j \\
			     \end{cases}
\end{equation}
$C_j$ represents the set of individuals originating from the $j^{th}$ component of the mixture model. Let $\phi = \left(p, \beta\right)$ denote the vector of model parameters, where $p = \left(p_1, p_2, ..., p_J\right)$ is the vector of mixing proportions, and $\beta = \left(\beta_1,  \beta_2, ..., \beta_J\right)$ is the regression coefficients vector. 
 The EM algorithm follows a two-step process to derive 
 the  estimating equations  using the log-likelihood $l(\phi |Y,Z,X)$ of the incomplete data:

\begin{equation}
		 	l(\phi  |Y,Z,X)  = \sum_{i=1}^{n} \sum_{j=1}^{J} z_{ij} \times log\left( p_{ij}\right)  + \sum \limits_{i=1}^{n} \sum \limits_{j=1}^{J} z_{ij} \times 	log\left(\frac{P_o(y_i,\lambda_{ij})}{\sum \limits_{k=0}^{\tau}P_o(k,\lambda_{ij})}\right).
\end{equation}  
 The first step, or E-step, computes the expected value of the incomplete data log-likelihood given the observed data and [$\phi = \phi^{(0)}$], where $\phi^{(0)} $ is a value of $\phi $ in the parameters space. This expected value  $Q\left(\phi | \phi^{(0)}\right)$ is maximized with respect to $\phi$ in the second step. For the details on these two steps we have:  
    \begin{enumerate}
	 	
	 	\item Given the observed data $\left(Y,X\right)$ and $\phi = \phi^{(0)}$, the expected  log-likelihood $Q\left(\phi | \phi^{(0)}\right)$ is expressed as follows:   
	 	\begin{eqnarray*}
	 		Q(\phi | \phi^{(0)}) & = & 
	 		E(l(P,\beta |Y,Z,X)|Y,X, \phi = \phi^{(0)})  \\ &=&
	 		\sum_{i=1}^{n}\sum_{j=1}^{J}E(z_{ij}|Y,W,\phi = \phi^{(0)} ) \times log(p_j) \\ &+&
	 		\sum_{i=1}^{n} \sum_{j=1}^{J}E(z_{ij}|Y,W,\phi = \phi^{(0)} ) \times log \left( \frac{ P_o(y_i,\lambda_{ij})}{\sum \limits_{k=0}^{\tau}P_o(k,\lambda_{ij})}\right) \\ &=&
	 		\sum_{i=1}^{n}\sum_{j=1}^{J}\alpha_{ij}^{(0)} \times log(p_j)+\sum_{i=1}^{n} \sum_{j=1}^{J}\alpha_{ij}^{(0)} \times log \left(\frac{ P_o(y_i,\lambda_{ij})}{\sum \limits_{k=0}^{\tau}P_o(k,\lambda_{ij})}\right)
	 	\end{eqnarray*}  
	 	with 
	 	\begin{eqnarray*}
	 		\alpha_{ij}^{(0)} &=&
	 		E(z_{ij} | Y,X,\phi = \phi^{(0)}) \\ &=&
	 		P(z_{ij} = 1 | Y,X,\phi = \phi^{(0)}) \\ &=&
	 		\frac{P(Y = y_i , z_{ij} = 1 | X,\phi = \phi^{(0)})}{P(Y = y_i | X, \phi = \phi^{(0)})}. \\   
	 	\end{eqnarray*}
	 	
	 	Therefore,
	 	
	 \begin{equation}
	 \alpha_{ij}^{(0)} =
	 \frac{p_j \times \frac{P_o(y_i,\lambda_{ij})}{\sum \limits_{k=1}^{\tau}P_o(k,\lambda)}}{\sum \limits_{j=1}^{J}p_j \times \frac{P_o(y_i,\lambda_{ij})}{\sum \limits_{k=1}^{\tau}P_o(k,\lambda)}}. \label{eq12}   
	 \end{equation}
	 	
	 $\alpha_{ij}^{(0)}$ is the posterior probability that  $y_i$ was generated from the j$^{th}$ component of the mixture, given $\phi =\phi^{(0)}$ and the observed data.
	 
	 \smallskip 

	 \item The maximization of $Q\left(\phi | \phi^{(0)} \right)$ in the M-step  is done through the Lagrangian,
	 	\begin{equation}
	 		Q^{'} (\phi | \phi^{(0)}) = Q(\phi | \phi^{(0)}) - \lambda(\sum_{j=1}^{J}p_j - 1) \label{eq13}.
	 	\end{equation}
	 	 Thus, the solution $ \hat{\phi}= (\hat{p},\hat{\beta})$  satisfies the first-order conditions
	 	\begin{eqnarray}
	 		\frac{\partial  Q^{'}(\phi | \phi^{(0)}) }{\partial p_j} &=& \sum_{i=1}^{n} \left( \frac{\alpha_{ij}^{(0)}}{p_j}-\lambda\right)  = 0,  \quad j \in  1,2,...,J-1  \label{eq14} \\
	 		\frac{\partial  Q^{'}(\phi | \phi^{(0)})}{\partial \beta_j}& =& \sum_{i=1}^{n} \alpha_{ij}^{(0)} \frac{\partial }{\partial \beta_{j}} log \left(\frac{ P_o(y_i,\lambda_{ij})}{\sum \limits_{k=0}^{\tau}P_o(k,\lambda_{ij})}\right)  = 0 , \quad j \in 1,...J  \label{eq15}
	 	\end{eqnarray}
	 	
	 	Equation (\ref{eq14}) implies
	 	
	 	\begin{equation}\label{eq16}
	 		\hat{p_j} =  \frac{\sum_{i= 1}^{n}\alpha_{ij}^{(0)}}{N}; \quad \forall j \in 1,...,J-1
	 	\end{equation}
	 	
	 	Solving equations \eqref{eq15} and \eqref{eq16} yields a solution to the likelihood optimization problem at each iteration of the EM algorithm.  As there is no closed form solution, a quasi-Newton method as proposed by~\cite{nash2018compact} is implemented. 
	 \end{enumerate}

    \subsection{ Properties of the estimators}\label{subsec4}

Under the identifiability of the regression model proposed above,  properties of the maximum likelihood estimator $\hat{\phi} = \left(\hat{p},\hat{\beta}\right)$   such as consistency and asymptotic normality can be established.  These  properties are stated in theorem~\ref{TheoAsymp} under regularity conditions of the model. 
 
    \begin{theorem} \label{TheoAsymp}
	 Assume that the regression model described in equations (\ref{eq5}) and (\ref{eq6}) is identifiable and that the number of components in the mixture is known. Then, under some  regularity conditions:
	 		\begin{itemize}
	 			\item $\hat{\phi}$ is consistent for estimating $\phi$,
	 			\item $ \sqrt{n}\left(\hat{\phi} - \phi\right)$ is asymptotically              normal, with a mean of zero and variance equal to the inverse of the  Fisher  information ($I(\phi)$) of the model.
	 		\end{itemize}
	 \end{theorem}
The regularity conditions  mentioned in this theorem are enumerated by ~\cite{lehmann2006theory}.   An approximation of the Fisher information is used to compute the variance in applications. This  approximation is based on the following expression: 
    \begin{equation}
		I_n\left(\hat{\phi}\right) = \left[  E_{Z|\phi=\hat{\phi}}\left(\frac{\partial l(\phi|Y,Z,X) }{\partial \phi}\right)   E_{Z|\phi=\hat{\phi}}\left(\frac{\partial l(\phi|Y,Z,X) }{\partial \phi}\right)^t \right]. \label{eq17}
    \end{equation}
	Although this definition of the Fisher information is less commonly used in the literature, it simplifies the computation  for mixture models.  It helps to avoid the computational drawbacks associated with the classical Fisher information approximation, defined as the negative expectation of the model's Hessian matrix. \cite{delattre2023estimatingfisherinformationmatrix} studied the properties and other characteristics of the Fisher information in equation~\eqref{eq17} and the reader is referred to their works for more information.

    \subsection{Algorithm and method to find an initial value}\label{subsec5}
    
    The EM algorithms alternate  between two global steps until convergence to a local maximum of the likelihood function.  The proposed algorithm for parameter estimation in the mixture of right-truncated Poisson mixture model proceeds as  follows: 
    
    \medskip

    \textit{\bf Step 1.}   Select the initial values
    $p^{(0)} =(p_1^{(0)},p_2^{(0)},...,p_J^{(0)}) $ and  
    $\beta^{(0)} = (\beta_1^{(0)},\beta_2^{(0)},...,\beta_J^{(0)} $)  as the starting point for the algorithm, and choose  two thresholds $\epsilon_1$ and $\epsilon_2$ which must remain fixed during iterations.

    \medskip

    \textit{\bf Step 2.} Determine the posterior probabilities $\alpha_{i}^{(0)} = (\alpha_{i1}^{(0)},\alpha_{i2}^{(0)},...,\alpha_{iJ}^{(0)})$ for $i \in \left\{1,2,...,n\right\}$ using equation \eqref{eq12} and  the vector $p^{(1)}$ of mixing weights using equation \eqref{eq16}. Solve equation \eqref{eq15} by using a quasi-Newton method  to determine the value $\beta^{(1)}$  of $\beta$ at iteration  $(1)$.

    \medskip

    \textit{\bf Step 3} If at least one of the following three conditions is met, set $\beta^{(0)} = \beta^{(1)}$ and $p^{(0)} = p^{(1)}$ and return to step 2. \\
    \smallskip
    \textbullet \quad $max( |  p^{(1)}-p^{(0)} |)  \geq \epsilon_1$ \\
    \textbullet \quad $max(|\beta^{(1)} -\beta^{(0)}| ) \geq \epsilon_1$ \\
    \textbullet \quad $l(p^{(1)},\beta^{(1)}|Y,X,Z)-l(p^{(0)},\beta^{(0)}|Y,X,Z)  \geq \epsilon_2$. \\
Otherwise, $p^{(1)}$ and $\beta^{(1)}$ are maximum likelihood estimators of the parameters $p$ and $\beta$. 
    \medskip

   \cite{wu1983convergence} and \cite{mclachlan2007algorithm}  addressed the  convergence issues of the EM algorithm for mixture models. According to their findings, the continuity of $Q\left(\phi|\phi^{(0)}\right)$ and its first derivative at $\phi$ and $\phi^{(0)}$ guarantee the convergence of the EM algorithm to a local maximum, global maximum, or saddle point of the observed likelihood function $l(p,\beta|Y,X).$ The type of convergence is strongly influenced by the  starting value of the algorithm. Noting that EM-type algorithms do not guarantee convergence to the global maximum, one has to select “coherent” initial values. We suggest a method that  provides an initial value to start the EM algorithm proposed here.  
   
    Let us assume that there are $J$ components in the mixture. The first step of the procedure  involves classifying the individuals into $J$ clusters based on the covariates matrix $X$ and using the k-means procedure by  ~\citep{macqueen1967some}.  In the second step, we use the data subset $\left(y_j,X_j\right)$ related to the j$^{th}$ cluster to estimate the coefficients of  regression  based on the RTPR model  $($see subsection~\ref{subsec1}$).$ We repeat this for all $J$ clusters and obtain a matrix of coefficients which serves as initial value $\beta^{(0)}$ for $\beta.$ The vector of proportions of individuals in each of the $J$ clusters from the  k-means procedure determines the initial value $p^{(0)}$ for the mixing weights $p.$  The vector $(p^{(0)},\beta^{(0)} )$ provides  a starting value for the EM algorithm.
    
    \subsection{Selecting the number of components}\label{subsec6}
    
    The number of components $(J)$  in a mixture model is generally unknown and  must be specified before applying the model to the data.  Choosing the number of components in a mixture model is a model selection problem in the same way as the significance of the regression coefficients.  
    
    Wald and likelihood ratio tests can be used to evaluate the significance of the regression coefficients when the true value of the number of components in the mixture is known. Standard likelihood ratio tests, however, cannot be used to determine the number of mixture components or to evaluate the significance of the regression coefficients when $J$ is unknown. This is because the mixing weights may lie near the boundary of the parameter space when the true value of the number of components is less than the fitted number of components.
    
    According to \cite{chernoff1954distribution} and \cite{self1987asymptotic}, the Chi-square approximation of the likelihood ratio test does not meet regularity requirements  in these cases. \cite{mclachlan1988mixture} proposed several different approaches to choose the number of components in a mixture model. For mixture component number selection using likelihood ratio tests, interested readers are  referred to \cite{ghosh1984asymptotic}, \cite{chen2009hypothesis}, and \cite{chen2001modified}.
    
     Here, we consider the  penalized likelihood criteria to select the number of components. The Akaike Information Criterion (AIC) and the Bayesian Information Criterion (BIC), proposed respectively by \cite{akaike1998information} and \cite{schwarz1978estimating} are two widely used criteria for model selection using penalized likelihood. 
	 AIC is based on decision theory and arises from maximizing an expected log-likelihood. This log-likelihood is penalizes by the number of parameters to prevent overfitting. BIC introduces a prior distribution over the parameter space. It penalizes model complexity more severely than AIC, especially when the number of observations is large. This characteristic makes BIC a more conservative criterion, often favoring simpler and more parsimonious models. In practice, this means that BIC tends to avoid overfitting more effectively than AIC, especially with large datasets. The use of information criteria in determining the number of components in a mixture model has been covered in  \cite{mclachlan1988mixture}. In this work,  AIC and BIC  are defined as follows : 
	\smallskip 
	\begin{itemize}
		\item  \( AIC =  - 2 \times l(\hat{p}, \hat{\beta} | Y, X) + 2 \times (J - 1 + J \times K) \) 
		\item  \( BIC = -2 \times l(\hat{p}, \hat{\beta} | Y, X) +   (J - 1 + J \times K) \times \log(n) \)
	\end{itemize}
	\smallskip 
    where $ l(\hat{\beta}, \hat{p} | Y, X) $ is the log-likelihood of the observed data at the estimated values $ (\hat{p}, \hat{\beta}) $, $ K $ the number of covariates, and $ n $ the sample size. Between two models, we select the one with the smallest values of AIC or BIC. Note that there are two steps involved in choosing the number of mixture model components  using AIC and BIC. First, we find the value of $J$ that minimizes AIC or BIC in the regression model with all covariates included. In the second step, either the $J$-components model with the lowest AIC or BIC is selected, or the model with $J$ components that only includes significant covariates is chosen. 

 From the simulation results presented in Table \ref{tab5} estimating the performances of information criteria in selecting the correct number of components, the BIC is extremely reliable, especially for large sample sizes. It always correctly identifies the number of mixture components. In contrast, AIC loses reliability as the sample size increases, tending to overestimate the number of model components and  making it less reliable compared to BIC. However, for small sample sizes, AIC is more reliable. For small sample sizes, BIC tends to underestimate the number of model components, whereas AIC identifies it with better accuracy. In conclusion, although both criteria are useful for identifying the correct number of components, BIC proves to be more reliable, especially for larger sample sizes. Therefore, we recommend using BIC for large sample sizes $(\geq 250)$ and AIC when the sample size is small. Note that these conclusions pertain to Configuration 1 of the simulation study, where the mean parameters $(\lambda )$ in the components are sufficiently separated.

 \subsection{Overdispersion}\label{subsec7}

In subsections~\ref{subsec1} and~\ref{subsec2}, we present two right-truncated regression models for underdispersed and overdispersed  truncated data respectively.  When analysing real-world data, it is not possible to know with certainty whether a data set is underdispersed, overdispersed or equidispersed. A simple way to check  for overdispersion is to compare the empirical mean to the empirical variance of the observed data without truncation.  When truncation is clearly identified or can be assumed in data,
~\cite{gurmu1992overdispersion} proposed a procedure 
to test for overdispersion in this data. We recommend using this procedure in real data analysis.   This test evaluates the null hypothesis of no overdispersion against the alternative hypothesis of overdispersion by considering a right-truncated negative binomial distribution for the count variable. The test statistic  has  a standard normal distribution under the null hypothesis. The reader is referred to~\cite{gurmu1992overdispersion}  for further details on  this testing approach.

 \section{Simulations study}\label{sec4}

    \subsection{Design}\label{subsec8}
 A Monte Carlo simulation was carried out to evaluate the performances of the MLE of parameters in the right-truncated Poisson mixture model  when the sample size is finite. Four configurations $($\textbf{Configs}$)$ are considered, as  listed in Table~\ref{tab1}, and for each of  them, we generate   $N = 100$ samples of size $n \in \left\{100,300,500,1000\right\}$ from a mixture of $J = 2$ components model. Several  values of   parameter vector $\left(p,\beta\right)$ are investigated to cover a wide range of scenarios. 
    \begin{table}[h!]
        \centering
        \caption{Configurations and the corresponding  parameter values}\label{tab1}
        \begin{tabular*}{\textwidth}{@{\extracolsep\fill}cccccc}
        \toprule%
        \textbf{Configs} & \textbf{$p_1$} & \textbf{$\beta_1$} & \textbf{$\beta_2$} & \textbf{$\tau$} & \textbf{Covariates} \\ 
        \midrule
        \textbf{1} & 0.3 & $(-1.2, 0.1)$ & $(1.5, -0.01)$ & 5 & $x_i = (1, v_i)$ with $v_i \sim \text{Uniform}(0, 20)$ \\ 
        \textbf{2} & 0.4 & $(-1.2, 0.1)$ & $(1.5, -0.01)$ & 5 & $x_i = (1, v_i)$ with $v_i \sim \text{Uniform}(0, 20)$ \\ 
        \textbf{3} & 0.1 & $(0.7, -0.3, 0.4)$ & $(1.58, -0.1, 0.3)$ & 6 & \text{ \makecell {$x_i = (1, v_i, u_i)$ with $v_i \sim \text{Uniform}(1, 3)$ \\ and $u_i \sim \text{Uniform}(0, 1)$}}\\ 
        \textbf{4} & 0.3 & $(0.3, 0.14)$ & $(0.9, -0.2)$ & 5 & $x_i = (1, v_i)$ with $v_i \sim \text{Normal}(3, 0.9)$ \\ 
        \toprule%
        \end{tabular*}
    \end{table} 

    The first configuration illustrates a case where there is a significant difference between the means of the truncated Poisson distributions in the two  components of the model. In particular, the average value of the count variable realizations in the first component is roughly equal to 1, whereas it is roughly equal to 4 in the second. In the second configuration, we investigate how the estimation method behaves when the components are distributed almost similarly. The third configuration evaluates the method in the case of imbalanced subpopulations, i.e. one of the mixture weights  is very small compared to the other, or when a mixing weight is close to the boundaries of its space. The last configuration represents a poor separation between the supports of the Poisson distributions in each component. In particular, the first component's mean is roughly 2.06, and the second component's mean is roughly 1.3. We report  the mean, the relative mean squared error $(ReMSE)$ and the relative bias $(RBias)$   to assess the performances of the MLE.  For a component $\alpha$ of the vector $\phi$ of  parameters, the $(ReMSE)$ and the  $(RBias)$ are defined as : 
	$$	ReMSE(\alpha)=\sum_{i=1}^{N}\frac{1}{n\alpha^2}(\hat{\alpha_i}-\alpha)^{2}, \mbox{ and }  RBias(\alpha)=\sum_{i=1}^{N}\frac{1}{n\alpha}(\hat{\alpha_i}-\alpha)$$
	where $\hat{\alpha_i}$ is the estimator of  $\alpha$ for the i$^{th}$ sample , $i \in \left\{1,2,...,N\right\}$.

    \subsection{ Results}\label{subsec10}
    
    \begin{sidewaystable}
    	\centering
    	\scriptsize
    	\caption{Simulation results for each configuration}
    	\begin{tabular}{cccccccccccccc}
    		\midrule
    		\multicolumn{14}{c}{\bf Configuration 1} \\
    		\midrule
    		&& \multicolumn{3}{c}{n = 100} & \multicolumn{3}{c}{n = 300}&\multicolumn{3}{c}{n = 500} 	& \multicolumn{3}{c}{n = 1000} \\
    		\cmidrule(lr){3-5} \cmidrule(lr){6-8} \cmidrule(lr){9-11} \cmidrule(lr){12-14}
    		Parameters & Value & Mean & ReMSE & RBias & Mean & ReMSE & RBias & Mean & ReMSE & RBias & Mean & ReMSE & RBias \\
    		\midrule
    		$p_1$ & 0.3 & 0.3000 & 0.1309 & 0.0117 & 0.3076 & 0.0937 & 0.0254 & 0.3156 & 0.0491 & 0.0519 & 0.3094 & 0.0277 & 0.0314 \\
    		$p_2$ & 0.7 & 0.7000 & 0.0240 & -0.0050 & 0.6924 & 0.0172 & -0.0109 & 0.6844 & 0.0090 & -0.0222 & 0.6906 & 0.0051 & -0.0134 \\
    		$\beta_{11}$ & -1.2 & -1.9203 & 17.3757 & 0.6003 & -1.7814 & 4.3743 & 0.4845 & -1.2617 & 0.3229 & 0.0514 & -1.2192 & 0.1288 & 0.0160 \\
    		$\beta_{12}$ & 0.1 & 0.1368 & 12.1713 & 0.3677 & 0.1456 & 2.5939 & 0.4555 & 0.1072 & 0.1811 & 0.0717 & 0.1022 & 0.0719 & 0.0224 \\
    		$\beta_{21}$ & 1.5 & 1.4469 & 0.1916 & -0.0354 & 1.4740 & 0.0673 & -0.0174 & 1.5351 & 0.0127 & 0.0234 & 1.5137 & 0.0051 & 0.0092 \\
    		$\beta_{22}$ & -0.01 & -0.0120 & 17.1320 & 0.2006 & -0.0127 & 5.3681 & 0.2694 & -0.0120 & 1.4725 & 0.2043 & -0.0106 & 0.6347 & 0.0571 \\
    		
    		\midrule
    		
    		\multicolumn{14}{c}{\bf Configuration 2} \\
    		\midrule
    		&& \multicolumn{3}{c}{n = 100} & \multicolumn{3}{c}{n = 300}&\multicolumn{3}{c}{n = 500} 	& \multicolumn{3}{c}{n = 1000} \\
    		\cmidrule(lr){3-5} \cmidrule(lr){6-8} \cmidrule(lr){9-11} \cmidrule(lr){12-14}
    		Parameters & Value & Mean & ReMSE & RBias & Mean & ReMSE & RBias & Mean & ReMSE & RBias & Mean & ReMSE & RBias \\
    		\midrule
    		$p_1$ & 0.4 & 0.3661 & 0.0616 & -0.0848 & 0.3900 & 0.0266 & -0.0249 & 0.3985 & 0.0162 & -0.0037 & 0.4052 & 0.0076 & 0.0131 \\
    		$p_2$ & 0.6 & 0.6339 & 0.0274 & 0.0565 & 0.6100 & 0.0118 & 0.0166 &  0.6015 & 0.0072 & 0.0025 & 0.5948 & 0.0034 & -0.0087 \\
    		$\beta_{11}$ &-1.2& -1.0869 & 7.0320 & -0.0942 & -1.4785 & 4.8376 & 0.2321 & -1.1526 & 0.3595 & -0.0395 & -1.2127 & 0.0531 & 0.0106 \\
    		$\beta_{12}$ &0.1& 0.0896 & 5.6949 & -0.1036 & 0.1263 & 2.6441 & 0.2630 & 0.1001 & 0.1533 & 0.0010 & 0.1019 & 0.0320 & 0.0188 \\
    		$\beta_{21}$ &1.5& 0.8602 & 0.6258 & -0.4265 & 1.3245 & 0.1862 & -0.1170 & 1.4701 & 0.0641 & -0.0199 & 1.5050 & 0.0048 & 0.0033 \\
    		$\beta_{22}$ &-0.01& 0.0153 & 30.4755 & -2.5268 & -0.0070 & 9.1635 & -0.2989 & -0.010 &  3.1846 & 0.0349 & -0.0105 & 0.6789 & 0.0471 \\
    		\midrule
    		
    		\multicolumn{14}{c}{\bf Configuration 3} \\
    		\midrule
    		&& \multicolumn{3}{c}{n = 100} & \multicolumn{3}{c}{n = 300}&\multicolumn{3}{c}{n = 500} 	& \multicolumn{3}{c}{n = 1000} \\
    		\cmidrule(lr){3-5} \cmidrule(lr){6-8} \cmidrule(lr){9-11} \cmidrule(lr){12-14}
    		Parameters & Value & Mean & ReMSE & RBias & Mean & ReMSE & RBias & Mean & ReMSE & RBias & Mean & ReMSE & RBias \\
    		
    		\midrule
    		
    		$p_1$ & 0.1&  0.2460 & 4.0202 & 1.4602 & 0.2620 & 5.4281 & 1.6202 & 0.2215 & 4.1292 & 1.2154 & 0.2083 & 3.4513 & 1.0827 \\
    		$p_2$ & 0.9 & 0.7540 & 0.0496 & -0.1622 & 0.7380 & 0.0670 & -0.1800 & 0.7785 & 0.0510 & -0.1350 & 0.7917 & 0.0426 & -0.1203 \\
    		$\beta_{11}$ & 0.7& 1.1625 & 34.7115 & 0.6608 & 1.3042 & 13.2702 & 0.8632 & 1.1630 & 15.6436 & 0.6614 & 0.8882 & 2.9695 & 0.2689 \\
    		$\beta_{12}$ & -0.3& -0.2325 & 38.5395 & -0.2249 & -0.4043 & 13.9403 & 0.3475 &-0.6007 & 27.0181 & 1.0025 & -0.2575 & 2.1946 & -0.1417 \\
    		$\beta_{13}$ & 0.4 &  0.5589 & 92.0230 & 0.3974 & 0.7152 & 40.6894 & 0.7880 &1.0364 & 58.0329 & 1.5910 & 0.4054 & 1.9563 & 0.0135 \\
    		$\beta_{21}$ &1.58 & 1.5069 & 0.1086 & -0.0462 & 1.5110 & 0.0703 & -0.0437 & 1.5312 & 0.0476 & -0.0309 & 1.5730 & 0.0230 & -0.0044 \\
    		$\beta_{22}$ & -0.1& -0.1013 & 4.2703 & 0.0129 & -0.0949 & 2.9571 & -0.0506 & -0.0835 & 1.8198 & -0.1649 & -0.0896 & 1.2711 & -0.1041 \\
    		$\beta_{23}$ & 0.3 & 0.3310 & 2.8508 & 0.1034 & 0.2693 & 1.2288 & -0.1024 & 0.2757 & 0.9692 & -0.0808 & 0.2836 & 0.4032 & -0.0546 \\
    		
    		\midrule
    		
    		\multicolumn{14}{c}{\bf Configuration 4} \\
    		\midrule
    		&& \multicolumn{3}{c}{n = 100} & \multicolumn{3}{c}{n = 300}&\multicolumn{3}{c}{n = 500} 	& \multicolumn{3}{c}{n = 1000} \\
    		\cmidrule(lr){3-5} \cmidrule(lr){6-8} \cmidrule(lr){9-11} \cmidrule(lr){12-14}
    		Parameters & Value & Mean & ReMSE & RBias & Mean & ReMSE & RBias & Mean & ReMSE & RBias & Mean & ReMSE & RBias \\
    		
    		\midrule
    		
    		$p_1$ & 0.3 & 0.2663 & 0.2728 & -0.1124 & 0.3392 & 0.2820 & 0.1308 & 0.3801 & 0.2857 & 0.2671 & 0.4216 & 0.3388 & 0.4052 \\
    		$p_2$ &0.7 & 0.7337 & 0.0501 & 0.0482 & 0.6608 & 0.0518 & -0.0561 & 0.6199  & 0.0525 & -0.1145 & 0.5784 & 0.0622 & -0.1737 \\
    		$\beta_{11}$ & 0.3 &  0.0562 & 207.7860 & -0.8126 & 0.6091 & 47.1369 & 1.0302 & 0.6381 & 29.9562 & 1.1271 & 0.4099 & 19.4003 & 0.3663 \\
    		$\beta_{12}$ &0.14&  -0.0132 & 93.7576 & -1.0942 & -0.1497 & 44.2803 & -2.0695 & -0.1302 & 32.2163 & -1.9300 & 0.0080 & 9.0614 & -0.9425 \\
    		$\beta_{21}$ & 0.9 & 0.6727 & 0.5595 & -0.2526 & 0.7355 & 0.2621 & -0.1828 & 0.6963 & 0.2124 & -0.2263 & 0.7281 & 0.2276 & -0.1910 \\
    		$\beta_{22}$ & -0.2 & -0.0854 & 1.4599 & -0.5732 & -0.0986 & 0.9010 & -0.5071 & -0.0996 & 0.8316 & -0.5020 & -0.1031 & 0.9240 & -0.4843 \\
    		\bottomrule
    	\end{tabular}
    \end{sidewaystable}
    
    \begin{table}[!h]
    	\centering
    	\footnotesize
    	\caption{Number of replicates where the selection criterion chose a right-truncated Poisson mixture model with the specified number of components when data of size \textit{n} was generated according to Configuration 1} \label{tab5}
    	\begin{tabular}{rrrrrrrrr}
    		\toprule
    		& \multicolumn{2}{c}{n = 100} & \multicolumn{2}{c}{n = 300} & \multicolumn{2}{c}{n = 500} & \multicolumn{2}{c}{n = 1000} \\
    		\cmidrule(lr){2-3} \cmidrule(lr){4-5} \cmidrule(lr){6-7} \cmidrule(lr){8-9}
    		{\makecell{Number of \\ components}} & AIC & BIC & AIC & BIC & AIC & BIC & AIC & BIC \\
    		\midrule
    		1 & 2 & 24 & 0 & 0 & 0 & 0 & 0 & 0 \\
    		\bf 2 & \bf 92 & \bf 75 & \bf 95 &\bf  100 & \bf 88 & \bf 100 & \bf 90 & \bf 100 \\
    		3 & 6 & 1 & 5 & 0 & 12 & 0 & 10 & 0 \\
    		\midrule
    		Total & 100 & 100 & 100 & 100 & 100 & 100 & 100 & 100 \\
    		\bottomrule
    	\end{tabular}
    \end{table}

    As  expected, both  RBias and  ReMSE decrease when the sample size increases. This is observed in all configurations, indicating that the MLE is suitable to estimate the model parameters in these contexts. The mean of the estimated regression coefficients for Configuration 1 is relatively close to the true parameter values, with a low  ReMSE expressing the accuracy of the method in this configuration.  Comparables conclusions are made for the results from Configuration 2. 
    
    Estimates of the mixture weights and regression coefficients for Configuration 3 are also accurate, but they require a larger sample size than the two previous configurations, especially for estimating the weights. When a parameter is close to the border of its membership space, the regularity conditions of the model tend to be violated, as the MLE can be unstable in this case, the results are somewhat anticipated.
          
      Results from Configuration 4 show that the method loses accuracy for a heterogeneous  population with components that are not well separated. In other words, the mean counts in the mixture components are close in value.  Also, we observe that the variances are typically less than their means for samples from this configuration.  The BIC suggests a unicomponent  regression rather than a mixture model, in this case. Consequently, we recommend  using the unicomponent right-truncated count regression wherever  variance is less than the mean. The overdispersion test can be  very helpful in directing  model selection in such a situation.

    \section{An application to real data}\label{sec5}

The main aim of this paper is to propose a regression method for seemingly right-truncated count data.  However, providing an example of data analysis using the proposed  method highlights its real-world applications. The method is applied to investigate the factors influencing adherence to prescribed  preventive measures  in Parakou, Benin, during the COVID-19 pandemic. Data are from a general survey  conducted by ODeSPoL $($Observatoire Démographique et Statistique des Populations Locales$)$ during  the COVID-19 period in Parakou, a city located in northern Benin  in  West Africa. They recorded a number of useful variables, including gender, internet access, monthly spending, education level, age, household size, and the number of  preventive measures adhered to.  A description of these variables is given in Table~\ref{tab2}.

    \begin{table}[h]
    	\caption{Description of variables analyzed to explain adherence level } \label{tab2}
    	\begin{tabular}{ll}
    		\toprule
    		\textbf{Variables} & \textbf{N = 1,516} \\
    		\midrule
    		\textbf{Age} & (21.40, 5.60) \\
    		\textbf{Gender} & \\
    		\textbf{Male} & 694 (46\%) \\
    		\textbf{Female} & 822 (54\%) \\
    		\textbf{Household Size} & (4.05, 5.56) \\
    		\textbf{Internet Access} & \\
    		\textbf{Yes} & 865 (57\%) \\
    		\textbf{No} & 651 (43\%) \\
    		\textbf{Education Level} & \\
    		\textbf{Primary} & 225 (15\%) \\
    		\textbf{Secondary} & 706 (47\%) \\
    		\textbf{Higher} & 524 (35\%) \\
    		\textbf{None} & 61 (4.0\%) \\
    		\textbf{Monthly Household Expenditure} & \\
    		\textbf{ Less than 50000} & 566 (38\%) \\
    		\textbf{ From 50000 to 150000} & 827 (54\%) \\
    		\textbf{ greater than 150000} & 123  (8\%) \\
    		\bottomrule
    	\end{tabular}
    \end{table}

\begin{figure}[H]
		\centering  
            \includegraphics[width=0.9\textwidth]{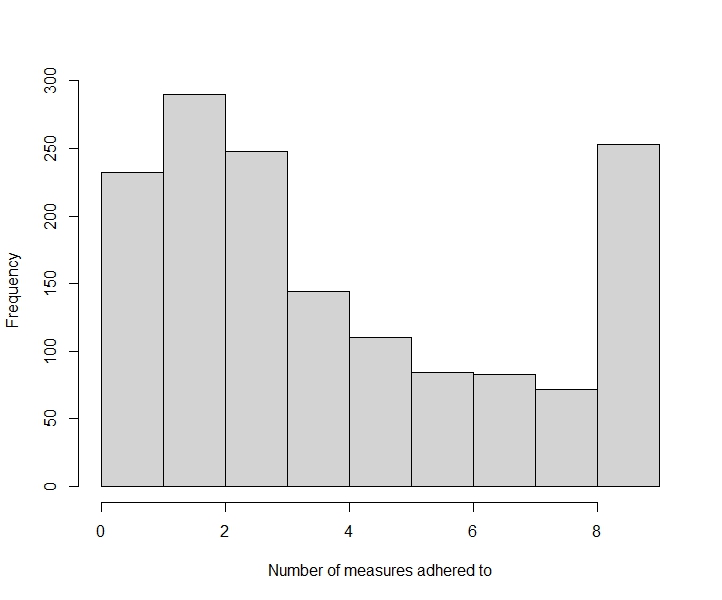}
		\caption{Histogram of the number of preventive measures adhered to by an individual} \label{fig1}
    \end{figure}

    Figure~\ref{fig1} presents the frequencies of the  number of preventative measures adhered to $($the response variable for the regression model$)$ by individuals during the COVID-19 period. As we can see, these frequencies do not correspond with a homogeneous Poisson distribution, and the difference $(8.23 - 4.32 = 3.90)$ between the variance and the mean indicates that the number of preventive measures adhered to seems overdispersed.  Also, the overdispersion test under  a truncation assumption \citep{gurmu1992overdispersion} has a p-value less than $2e^{-16}.$ This indicates that the number of preventive measures can be analysed using  a mixture of Poisson distributions.  Therefore, we propose a mixture of right-truncated Poisson distributions  to describe variations of these counts given covariates.

   Here, the truncation threshold is equal to the number of barrier measures prescribed $(\tau=9).$ Both the $\tau$-right-truncated Poisson mixture model and the untruncated Poisson mixture model are applied to the number of preventive measures adhered to. We first apply these two regression models using all of the covariates available and increasing the number of mixture components.  For each number $J=1,2, \dots$ of components, AIC and BIC are computed and compared to their previous value. If at a value $j_0,$ the information criterion grows or does not significantly decrease,  then $j_0$ is considered as the selected number of  components in the mixture model. From the results in Table \ref{tab3}, this procedure suggests a mixture of two  $(J=2)$ components for both untruncated Poisson and $\tau$-right-truncated Poisson mixture models.  Backward selection is used to select  the significant regressors  for the final model. A description of this  selection method can be found in~\cite{cornillon2023regression}. Our application identifies the following four covariates out of the six presented in Table \ref{tab2}:   internet access, age, household size, and education level.  The  likelihood ratio test is used to assess the significance of each regressor in the model.

	\begin{table}[h]
		\caption{Number of Components and Information Criteria} 
		\label{tab3}
		\centering
		\begin{tabular}{@{}lllll@{}}
			\toprule
			& \multicolumn{2}{c}{\makecell{ \bf Mixed right-truncated \\ \bf Poisson regression}} & \multicolumn{2}{c}{\makecell{\bf Mixed Poisson \\ \bf regression}}  \\
			\midrule
			\textbf{\makecell {Number of\\ components}} & \textbf{AIC} & \textbf{BIC} & \textbf{AIC} & \textbf{BIC} \\ 
			\midrule
			1 & 7599.806 & 7653.044 & 7599.806 & 7653.044 \\
			2 & \bf 6579.549 & \bf 6691.349 & \bf 7134.889 & \bf 7246.689 \\
			3 & \bf 6568.691 & 6739.054 & 7142.887 & 7313.249 \\
			4 & 6586.587 & 6815.512 & 7160.039 & 7388.963 \\
			\bottomrule
		\end{tabular}
	\end{table}

\begin{figure}[h]
    \centering
    \includegraphics[width=0.9\textwidth]{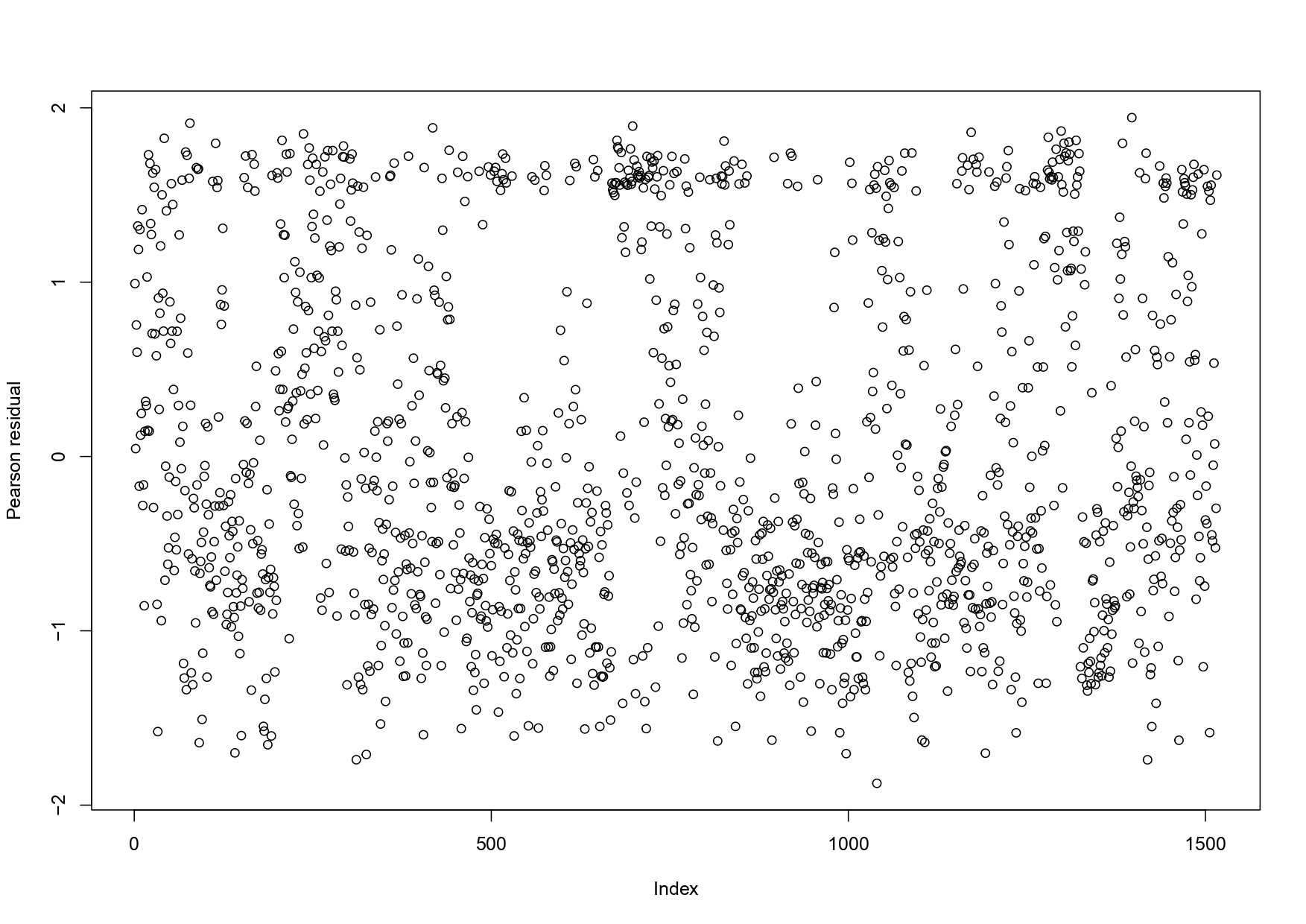}
    \caption{Individual Pearson Residuals} \label{fig2}
    \end{figure}

    \begin{table}[h]
    \caption{Results of the right-truncated Poisson mixture model fitted to the number of measures adhered to and its covariates}\label{tab4}%
    \begin{tabular}{@{}lcccc@{}}
    \toprule
        \multicolumn{1}{c}{ } & \multicolumn{3}{c}{Component 1: $\mathbf{p_1 = 0.26}$} \\
	\cmidrule(l{3pt}r{3pt}){2-5}
	\textbf{ } & \textbf{Coefficients} & \textbf{Standard Error} & \textbf{z} &              \textbf{P($\left|Z\right|\geq$ z)}\\
	\midrule
	\textcolor{black}{\textbf{Intercept}} & 2.4967 & 0.4921 & 5.0731 & 0.0000\\
	\textcolor{black}{\textbf{Age}} & 0.0456 & 0.0134 & 3.3940 & 0.0007\\
	\textcolor{black}{\textbf{Education Level-Primary}} & -0.0822 & 0.4721 & -0.1741 &        0.8617\\
	\textcolor{black}{\textbf{Education Level-Secondary}} & -0.6241 & 0.4465 & -1.3979         & 0.1621\\
	\textcolor{black}{\textbf{Education Level-Higher}} & -0.3854 & 0.4760 & -0.8096 &          0.4182\\
	\textcolor{black}{\textbf{Household Size}} & 0.0105 & 0.0311 & 0.3391 & 0.7345\\
	\textcolor{black}{\textbf{Internet Access-Yes}} & -0.0614 & 0.1682 & -0.3651 &             0.7150\\
	\midrule
	\addlinespace
	\multicolumn{1}{c}{ } & \multicolumn{3}{c}{Component 2: $\mathbf{p_2 = 0.74}$} \\
	\cmidrule(l{3pt}r{3pt}){2-5}
	\textbf{ } & \textbf{Coefficients} & \textbf{Standard Error} & \textbf{z} &        \textbf{P($\left|Z\right|\geq$ z)}\\
	\midrule
	\textcolor{black}{\textbf{Intercept}} & 0.4746 & 0.1481 & 3.2040 & 0.0014\\
	\textcolor{black}{\textbf{Age}} & 0.0013 & 0.0037 & 0.3489 & 0.7271\\
	\textcolor{black}{\textbf{Education Level-Primary}} & 0.2456 & 0.1260 & 1.9500 &           0.0512\\
	\textcolor{black}{\textbf{Education Level-Secondary}} & 0.4003 & 0.1190 & 3.3629 &         0.0008\\
        \textcolor{black}{\textbf{Education Level-Higher}} & 0.4600 & 0.1209 & 3.8056 & 0.0001\\
	\textcolor{black}{\textbf{Household Size}} & 0.0299 & 0.0088 & 3.4051 & 0.0007\\
	\textcolor{black}{\textbf{Internet Access-Yes}} & 0.1046 & 0.0501 & 2.0895 &               0.0367\\
	\midrule
	\addlinespace
	\bf AIC =  6571.799 &&&& \bf BIC = 6651.656 \\
    \end{tabular}
    \end{table}
   Table~\ref{tab4} summarises the estimation results of the two-component  truncated Poisson mixture model using the four significant variables. Results from the untruncated  two-component Poisson mixture model with the same variables  are presented in Table ~\ref{mixedP} in Appendix. Except for internet access, the variable selection methods agree in selecting the other variables  either  in the untruncated Poisson or  the right-truncated Poisson mixture model.  Figure \ref{fig2} shows the Pearson residual  for each observation in the right-truncated regression model. All residuals are inside  the interval $\left[-2, 2\right]$, indicating a good fit for the individual data. The Pearson goodness-of-fit  test also confirms a good fit to the data. 
   
   Both models adjusted a similar mixture of two components    model  such that in the first component $($with weights $0.26$ and $0.42),$ only age is significant at level $0.05.$ All of the other variables (except age)  contribute  significantly to  explain the variation in the number of measures adhered to in the second component. Despite this similar results,  the truncated model seems to offer a better fit to our data, which displays a known truncation level $($at $\tau = 9).$ We note that the older an individuals is in the first subpopulation, the more he adheres to the prescribed prevention measures. In the second subpopulation, a higher level of education, greater   access to internet and a larger household size are each associated with an increase in the number of measures adhered to. The variations in adherence seem contingent upon the subpopulation from which individuals originate.
        
    \section{Discussion}\label{sec6}

In the application prensented here, two mixture components are  identified through an effective selection procedure. In the first component,  increasing age correlates with an increase in the number of preventive measures; in the second component, education level and household size increase the number of preventive measures.  These results are consistent with those obtained by \cite{bouton2022transmission}.  Furthermore, internet access increases the number of preventive measures in the second component only.   This could be attributed to the fact that social media platforms raise awareness of COVID-19's risks due to its high mortality rate in the West. 
 
 The regression method we propose accounts for both heterogeneity and truncation, if present in the data under analysis.   The simulation results presented in section \ref{sec4} confirm the identifiability assumption  of the mixture of  right-truncated distributions used in this method.  When the truncation threshold $\tau$ is large, far beyond the range of observed data, the right-truncation method behaves similarly to the one proposed by ~\cite{wang1996mixed}. However,  when there is evidence of truncation in the data, the estimated regression coefficients  are affected and the truncated method is more suitable.  
 
    \section{Conclusion}\label{sec7}

   We investigated a right-truncated Poisson mixture model for count data analysis.  This regression method can  explain right-truncated count data while allowing regression coefficients to vary between components. An EM algorithm is proposed to compute the maximum likelihood estimators for  regression coefficients and mixing weights. The simulation results demonstrated that, the right-truncated mixture model is identifiable, and implementing the method in R statistical software is efficient. Simulations  also show that model selection  based on Bayesian Information Criterion  provides very good results. We 
   applied the model to real-world count data on adherence to preventative measures during COVID-19 in northern Benin. This application provided insights into the determinants of adherence. An interesting direction for further research would be extending the model to allow the mixing weights to depend  on covariates. Although the simulation results reassurance on the identifiability of the model, a theoretical foundation should also be established for our results.

	\newpage

\section{Appendices} \label{sec8}

\subsection{Proofs}\label{subsec8.1}

	\subsubsection{Proof of Theorem~\ref{theo1}} \label{subsubsec8.1.1}
	
	Let \( Y \) be a random variable taking values in the set \( \{0, 1, \ldots, \tau\} \), \( f_\lambda \)  the Poisson probability mass function and \( F_\lambda \) its distribution function such that the p.m.f. of   \( Y \) is given by

	\[
	P(Y = y) = \frac{f_\lambda(y)}{F_\lambda(\tau)} \quad
	\]
	with \( f_\lambda(y) = \frac{\lambda^y e^{-\lambda}}{y!} \) and \( F_\lambda(y) = \sum \limits_{k=0}^{\tau} f_\lambda(k) \).
	
	We have :
	\begin{eqnarray*}
		E(Y) &=&
		\sum \limits_{y=0}^{\tau} yP(Y=y) \\
		&=&  \sum \limits_{y=0} ^{\tau}y \times \frac{\frac{\lambda^{y}e^{-\lambda}}{y!}}{\sum \limits_{k=0}^{\tau}\frac{\lambda^{k}e^{-\lambda}}{k!}} \\
		&=& \frac{1}{F_\lambda(\tau)} \times \left(\sum \limits_{y=0}^{+\infty}y\frac{e^{-\lambda}\lambda^{y}}{y!} - \sum \limits_{y>\tau}y\frac{e^{-\lambda}\lambda^{y}}{y!}\right) \\
		&=& \frac{1}{F_\lambda(\tau)} \times \left(\lambda - \lambda \sum \limits_{y>\tau}\frac{e^{\lambda}\lambda^{y-1}}{(y-1)!}\right) \\
		&=& \frac{\lambda}{F_\lambda(\tau)} \times \left(1 - \sum \limits_{x>\tau-1} \frac{e^{\lambda}\lambda^{x}}{x!}\right)  \\
		&=& \frac{\lambda}{F_\lambda(\tau)} \sum \limits_{x=0}^{\tau-1} \frac{e^{\lambda}\lambda^x}{x!} \\
		&=& \lambda\frac{F_\lambda(\tau-1)}{F_\lambda(\tau)} \\
		E(X) &=& \lambda \left(1-\frac{f_{\lambda}(\tau)}{F_{\lambda}(\tau)}\right)
	\end{eqnarray*}
	
	We also have : 
	
	\begin{eqnarray*}
		E(Y^2) &=& 
		E(Y(Y-1))+E(Y) \\
		&=& \frac{1}{F_\lambda(\tau)}\left[\sum \limits _{y=0}^{\tau}y(y-1)\frac{\lambda^ye^{-\lambda}}{y!}\right] + E(Y) \\
		&=& \frac{1}{F_\lambda(\tau)} \left[\sum \limits_{y=0}^{+\infty}y(y-1)\frac{\lambda^ye^{-\lambda}}{y!} - \sum \limits_{y>\tau} y(y-1)\frac{\lambda^ye^{-\lambda}}{y!}\right] + E(Y) \\
		&=& \frac{1}{F_\lambda(\tau)}\left[\lambda^2 -  \lambda^2\sum \limits_{y>\tau}\frac{\lambda^{y-2}e^{-\lambda}}{(y-2)!} \right] + E(Y) \\
		&=& \frac{\lambda^2}{F_\lambda(\tau)} \left[1- \sum \limits _{x>\tau -2} \frac{\lambda^xe^{-\lambda}}{x!} \right] + E(Y) \\
		&=& \frac{\lambda^2}{F_\lambda(\tau)}\left[\sum \limits _{x=0}^{\tau-2} \frac{\lambda^xe^{-\lambda}}{x!}\right] + E(Y) \\
		&=& \lambda^2 \frac{F_\lambda(\tau-2)}{F_\lambda(\tau)} + \lambda\frac{F_\lambda(\tau-1)}{F_\lambda(\tau)} \\
		E(X^2) &=&  \lambda^2\left(1 - \frac{f_{\lambda}(\tau)+f_{\lambda}(\tau-1)}{F_{\lambda}(\tau)}\right) + \lambda \left(1-\frac{f_{\lambda}(\tau)}{F_{\lambda}(\tau)}\right) , \tau>=2 
	\end{eqnarray*}
	
	Then, 
	
	\begin{eqnarray*}
		V(Y) &=&
		E(Y^2) - \left[E(Y)\right]^2 \\
		V(Y) &=&   \lambda^2\left(1 - \frac{f_{\lambda}(\tau) + f_\lambda(\tau-1)}{F_{\lambda}(\tau)}\right) + \lambda \left(1-\frac{f_{\lambda}(\tau)}{F_{\lambda}(\tau)}\right) - \lambda^2 \left[1-\frac{f_\lambda(\tau)}{F_\lambda(\tau)} \right] ^ 2  , \tau>=2
	\end{eqnarray*}
	
	It follows that : 
	
	\begin{eqnarray*}
		V(X)-E(X)  &=& \lambda^2\left(1 - \frac{f_{\lambda}(\tau)+f_\lambda(\tau-1)}{F_{\lambda}(\tau)}\right) - \lambda^2\left[1-\frac{f_\lambda(\tau)}{F_\lambda(\tau)} \right] ^ 2  \\
		&=& \lambda^2 \left[ 1- \frac{f_\lambda(\tau)+f_\lambda(\tau-1)}{F_\lambda(\tau)}   \right] - \lambda^2 \left[ 1 - 2 \frac{f_\lambda(\tau)}{F_\lambda(\tau)} + \left( \frac{f_\lambda(\tau)}{F_\lambda(\tau)}\right) ^2\right] \\
		&=& -\lambda^2 \left[ \left( \frac{f_\lambda(\tau)}{F_\lambda(\tau)}\right) ^2 + \frac{f_\lambda(\tau-1) - f_\lambda(\tau)}{F(\tau)}   \right] 
	\end{eqnarray*}
			$V(X) -E(X) <0$  because for all $\lambda$ such that  {$\lambda \leq \tau$,}  we have $f_\lambda(\tau-1)-f_\lambda(\tau)  \geq 0$  
	
	\medskip

	\subsubsection{Expectation and Variance of the Discrete Mixture of Truncated Distributions}\label{subsubsec8.1.2}

	We use the same notations as in the previous section and consider the truncated Poisson mixture model in (4). We have:
	\[
	P(Y_i = y_i) = \sum_{j=1}^{J} p_j \times \frac{f_{\lambda_{ij}}(y_i)}{F_{\lambda_{ij}}(\tau)}
	\]
	
	Thus, 
	
	\begin{eqnarray*}
		E(Y_i) &=& E\left[E \left(Y_i|Z_i\right)\right]  \\
		&=& \sum \limits_{j=1}^{J} E(Y_i|Z_i=j) \times P(Z_i = j) \\
		E(Y_i) &=& \sum \limits_{j=1}^{J} p_j \times \lambda_{ij}\frac{F_{\lambda_{ij}}(\tau-1)}{F_{\lambda_{ij}}(\tau)} \\
		&\quad& car \quad E(Y_i|Z_i=j) = \lambda_{ij} \frac{F_{\lambda_{ij}}(\tau-1)}{F_{\lambda_{ij}}(\tau)}
	\end{eqnarray*} 
	
	We also have : 
	
	\begin{equation*}
		V(Y_i) 
		= E\left[Var(Y_i|Z_i)\right] + Var\left[E(Y_i|Z_i)\right] 
	\end{equation*}
	
	but , \begin{equation*}
		Var(Y_i | Z_i = j) = \lambda_{ij}^2 \frac{F_{\lambda_{ij}}(\tau-2)}{F_{\lambda_{ij}}(\tau)} + \lambda_{ij} \frac{F_{\lambda_{ij}}(\tau-1)}{F_{\lambda_{ij}}(\tau)} - \left[\lambda_{ij} \frac{F_{\lambda_{ij}}(\tau-1)}{F_{\lambda_{ij}}(\tau)}\right]^2
	\end{equation*}
	
	and \begin{equation*}
		E(Y_i|Z_i = j) = \lambda_{ij} \frac{F_{\lambda_{ij}}(\tau-1)}{F_{\lambda_{ij}}(\tau)}
	\end{equation*}
	
	Then , 
	
	\begin{equation*}
		Var(Y_i) =  \sum \limits_{j=1}^{J}p_j \lambda_{ij}^2\frac{F_{\lambda_{ij}}(\tau-2)}{F_{\lambda_{ij}}(\tau)}  + \sum \limits_{j=1}^J p_j\lambda_{ij}\frac{F_{\lambda_{ij}}(\tau-1)}{F_{\lambda_{ij}}(\tau)} - \left[ \sum \limits_{j=1}^Jp_j\lambda_{ij} \frac{F_{\lambda_{ij}}(\tau-1)}{F_{\lambda_{ij}}(\tau)}\right]^2 
	\end{equation*}
	\newpage
	
	\subsection{Tables}%\label{subsec8.2}
	
	\begin{table}[h!]
		\centering
		\footnotesize
		\caption{Results of the untruncated Poisson mixture model fitted to the number of measures adhered to and its covariates} \label{mixedP}
			\begin{tabular}{lrrrr}
				\toprule
				\multicolumn{1}{c}{ } & \multicolumn{3}{c}{Component 1: $\mathbf{p_1 = 0.42}$} \\
				\cmidrule(l{3pt}r{3pt}){2-5}
				\textbf{ } & \textbf{Coefficients} & \textbf{Standard Error} & \textbf{z} &  \textbf{P($\left|Z\right|\geq$ z)}\\
				\midrule
				\textcolor{black}{\textbf{Intercept}} & 1.7250 & 0.1647 & 10.4722 & $<$2e-16 \\
				\textcolor{black}{\textbf{Age}} & 0.0089 & 0.0043 & 2.0547 & 0.0399 \\
				\textcolor{black}{\textbf{Education Level-Primary}} & 0.0413 & 0.1479 & 0.2789 & 0.7803\\
				\textcolor{black}{\textbf{Education Level-Secondary}} & -0.0343 & 0.1357 & -0.2527 & 0.8005\\
				\textcolor{black}{\textbf{Education Level-Higher}} & -0.0538 & 0.1373 & -0.3915 & 0.6955\\
				\textcolor{black}{\textbf{Household Size}} & 0.0086 & 0.0110 & 0.7791 & 0.4359\\
				\midrule
				
				\multicolumn{1}{c}{ } & \multicolumn{3}{c}{Component 2: $\mathbf{p_2 = 0.58}$} \\
				\cmidrule(l{3pt}r{3pt}){2-5}
				\textbf{ } & \textbf{Coefficients} & \textbf{Standard Error} & \textbf{z} &  \textbf{P($\left|Z\right|\geq$ z)}\\
				\midrule
				\textcolor{black}{\textbf{Intercept}} & 0.3882 & 0.2070 & 1.8757 & 0.0607\\
				\textcolor{black}{\textbf{Age}} & -0.0008 & 0.0054 & -0.1537 & 0.8778\\
				\textcolor{black}{\textbf{Education Level-Primary}} & 0.1809 & 0.1793 & 1.0087 & 0.3131\\
				\textcolor{black}{\textbf{Education Level-Secondary}} & 0.4040 & 0.1572 & 2.5701 & 0.0102 \\
				\textcolor{black}{\textbf{Education Level-Higher}} & 0.4759 & 0.1570 & 3.0306 & 0.0024 \\
				\textcolor{black}{\textbf{Household size}} & 0.0327 & 0.0130 & 2.5207 & 0.0117 \\
				\bottomrule
				\addlinespace
				\bf AIC =  7125.896 &&&& \bf BIC = 7195.106 \\
		\end{tabular}
	\end{table}
	
	\newpage
    
\bibliography{main}

\begin{thebibliography}{}

\bibitem[Akaike, 1998]{akaike1998information}
Akaike, H. (1998).
\newblock Information theory and an extension of the maximum likelihood
  principle.
\newblock In {\em Selected Papers of Hirotugu Akaike}, pages 199--213.
  Springer, New York, USA.

\bibitem[Bouton et~al., 2022]{bouton2022transmission}
Bouton, C., Meziere, P., and Rat, C. (2022).
\newblock Transmission de la covid 19: Identification de facteurs associ{\'e}s
  {\`a} des comportements non conformes aux recommandations de pr{\'e}vention.
  une enqu{\^e}te par questionnaire aupr{\`e}s de 1004 professionnels.
\newblock In {\em Congr{\`e}s du Coll{\`e}ge National des G{\'e}n{\'e}ralistes
  Enseignants}.

\bibitem[Breslow, 1984]{breslow1984extra}
Breslow, N.~E. (1984).
\newblock Extra-poisson variation in log-linear models.
\newblock {\em Journal of the Royal Statistical Society: Series C (Applied
  Statistics)}, 33(1):38--44.

\bibitem[Chen et~al., 2001]{chen2001modified}
Chen, H., Chen, J., and Kalbfleisch, J.~D. (2001).
\newblock A modified likelihood ratio test for homogeneity in finite mixture
  models.
\newblock {\em Journal of the Royal Statistical Society Series B: Statistical
  Methodology}, 63(1):19--29.

\bibitem[Chen and Li, 2009]{chen2009hypothesis}
Chen, J. and Li, P. (2009).
\newblock Hypothesis test for normal mixture models: The em approach.

\bibitem[Chernoff, 1954]{chernoff1954distribution}
Chernoff, H. (1954).
\newblock On the distribution of the likelihood ratio.
\newblock {\em The Annals of Mathematical Statistics}, pages 573--578.

\bibitem[Cornillon et~al., 2023]{cornillon2023regression}
Cornillon, P.-A., Hengartner, N., Matzner-L{\o}ber, E., and Rouvi{\`e}re, L.
  (2023).
\newblock {\em R{\'e}gression avec R: 3{\`e}me {\'e}dition}.
\newblock EDP Sciences, Paris, France.

\bibitem[Delattre and Kuhn,
  2023]{delattre2023estimatingfisherinformationmatrix}
Delattre, M. and Kuhn, E. (2023).
\newblock Estimating fisher information matrix in latent variable models based
  on the score function.

\bibitem[Dempster et~al., 1977]{dempster1977maximum}
Dempster, A.~P., Laird, N.~M., and Rubin, D.~B. (1977).
\newblock Maximum likelihood from incomplete data via the em algorithm.
\newblock {\em Journal of the royal statistical society: series B
  (methodological)}, 39(1):1--22.

\bibitem[Follmann and Lambert, 1989]{follmann1989generalizing}
Follmann, D.~A. and Lambert, D. (1989).
\newblock Generalizing logistic regression by nonparametric mixing.
\newblock {\em Journal of the American Statistical Association},
  84(405):295--300.

\bibitem[Ghosh and Sen, 1984]{ghosh1984asymptotic}
Ghosh, J.~K. and Sen, P.~K. (1984).
\newblock On the asymptotic performance of the log likelihood ratio statistic
  for the mixture model and related results.
\newblock Technical report, North Carolina State University. Dept. of
  Statistics.

\bibitem[Gurmu and Trivedi, 1992]{gurmu1992overdispersion}
Gurmu, S. and Trivedi, P.~K. (1992).
\newblock Overdispersion tests for truncated poisson regression models.
\newblock {\em Journal of Econometrics}, 54(1-3):347--370.

\bibitem[Hausman et~al., 1984]{hausman1984econometric}
Hausman, J.~A., Hall, B.~H., and Griliches, Z. (1984).
\newblock Econometric models for count data with an application to the
  patents-r\&d relationship.

\bibitem[Hinde, 1982]{hinde1982compound}
Hinde, J. (1982).
\newblock Compound poisson regression models.
\newblock In {\em Glim 82: Proceedings of the international conference on
  generalised linear models}, pages 109--121. Springer.

\bibitem[Lee, 1986]{lee1986specification}
Lee, L.-F. (1986).
\newblock Specification test for poisson regression models.
\newblock {\em International Economic Review}, pages 689--706.

\bibitem[Lehmann and Casella, 2006]{lehmann2006theory}
Lehmann, E.~L. and Casella, G. (2006).
\newblock {\em Theory of Point Estimation}.
\newblock Springer Science \& Business Media, New York, USA.

\bibitem[MacQueen et~al., 1967]{macqueen1967some}
MacQueen, J. et~al. (1967).
\newblock Some methods for classification and analysis of multivariate
  observations.
\newblock In {\em Proceedings of the fifth Berkeley symposium on mathematical
  statistics and probability}, volume~1, pages 281--297. Oakland, CA, USA.

\bibitem[Manton et~al., 1981]{manton1981variance}
Manton, K.~G., Woodbury, M.~A., and Stallard, E. (1981).
\newblock A variance components approach to categorical data models with
  heterogenous cell populations: analysis of spatial gradients in lung cancer
  mortality rates in north carolina counties.
\newblock {\em Biometrics}, pages 259--269.

\bibitem[McCullagh and Nelder, 1989]{mccullagh1989generalized}
McCullagh, P. and Nelder, J. (1989).
\newblock Generalized linear models.

\bibitem[McLachlan and Basford, 1988]{mclachlan1988mixture}
McLachlan, G.~J. and Basford, K.~E. (1988).
\newblock {\em Mixture Models: Inference and Applications to Clustering},
  volume~38.
\newblock Marcel Dekker, New York, USA.

\bibitem[McLachlan and Krishnan, 2007]{mclachlan2007algorithm}
McLachlan, G.~J. and Krishnan, T. (2007).
\newblock {\em The EM algorithm and extensions}.
\newblock John Wiley \& Sons, New York.

\bibitem[Nash, 2018]{nash2018compact}
Nash, J.~C. (2018).
\newblock {\em Compact Numerical Methods for Computers: Linear Algebra and
  Function Minimisation}.
\newblock Routledge, London and New York.

\bibitem[Schall, 1991]{schall1991estimation}
Schall, R. (1991).
\newblock Estimation in generalized linear models with random effects.
\newblock {\em Biometrika}, 78(4):719--727.

\bibitem[Schwarz, 1978]{schwarz1978estimating}
Schwarz, G. (1978).
\newblock Estimating the dimension of a model.
\newblock {\em The annals of statistics}, pages 461--464.

\bibitem[Self and Liang, 1987]{self1987asymptotic}
Self, S.~G. and Liang, K.-Y. (1987).
\newblock Asymptotic properties of maximum likelihood estimators and likelihood
  ratio tests under nonstandard conditions.
\newblock {\em Journal of the American Statistical Association},
  82(398):605--610.

\bibitem[Wang et~al., 1996]{wang1996mixed}
Wang, P., Puterman, M.~L., Cockburn, I., and Le, N. (1996).
\newblock Mixed poisson regression models with covariate dependent rates.
\newblock {\em Biometrics}, pages 381--400.

\bibitem[Williams, 1982]{williams1982extra}
Williams, D.~A. (1982).
\newblock Extra-binomial variation in logistic linear models.
\newblock {\em Journal of the Royal Statistical Society: Series C (Applied
  Statistics)}, 31(2):144--148.

\bibitem[Wu, 1983]{wu1983convergence}
Wu, C.~J. (1983).
\newblock On the convergence properties of the em algorithm.
\newblock {\em The Annals of statistics}, pages 95--103.

\end{thebibliography}
\bibliographystyle{apalike}

\end{document}